\documentclass[prd,superscriptaddress,amsfonts,amssymb,amsmath,showpacs,twocolumn]{revtex4-2}
\usepackage{bm}
\usepackage{amsfonts}
\usepackage{latexsym}
\usepackage[latin1]{inputenc}
\usepackage{graphicx}
\usepackage{amsmath}
\usepackage{palatino}
\usepackage{mathpazo}
\usepackage{textcomp}
\usepackage{float}
\usepackage{booktabs}
\usepackage{dcolumn}
\usepackage{hyperref}
\hypersetup{colorlinks,citecolor=blue}
\usepackage{amsmath}
\usepackage{xcolor}
\usepackage{orcidlink}
\usepackage[caption=false]{subfig}
\usepackage{commath}
\def\jnl@style{\it}
\def\aaref@jnl#1{{\jnl@style#1}}

\def\aaref@jnl#1{{\jnl@style#1}}

\def\aj{\aaref@jnl{AJ}}                   
\def\apj{\aaref@jnl{ApJ}}                 
\def\apjl{\aaref@jnl{ApJ}}                
\def\apjs{\aaref@jnl{ApJS}}               
\def\apss{\aaref@jnl{Ap\&SS}}             
\def\aap{\aaref@jnl{A\&A}}                
\def\aapr{\aaref@jnl{A\&A~Rev.}}          
\def\aaps{\aaref@jnl{A\&AS}}              
\def\mnras{\aaref@jnl{Mon.~Not.~Roy.~Astron.~Soc.}}             
\def\prd{\aaref@jnl{Phys.~Rev.~D}}        
\def\prc{\aaref@jnl{Phys.~Rev.~C}}  
\def\prl{\aaref@jnl{Phys.~Rev.~Lett.}}    
\def\qjras{\aaref@jnl{QJRAS}}             
\def\skytel{\aaref@jnl{S\&T}}             
\def\ssr{\aaref@jnl{Space~Sci.~Rev.}}     
\def\zap{\aaref@jnl{ZAp}}                 
\def\nat{\aaref@jnl{Nature}}              
\def\aplett{\aaref@jnl{Astrophys.~Lett.}} 
\def\apspr{\aaref@jnl{Astrophys.~Space~Phys.~Res.}} 
\def\physrep{\aaref@jnl{Phys.~Rep.}}      
\def\physscr{\aaref@jnl{Phys.~Scr}}       
\def\commat{\aaref@jnl{Comm.~Math.~Phys.}}              
\def\science{\aaref@jnl{Science}}               
\def\cqg{\aaref@jnl{Classical Quant.~Grav.}}            
\def\jpcs{\aaref@jnl{JPCS}}                                     
\def\ijmpd{\aaref@jnl{Int.~J.~Mod.~Phys.~D}}                    
\def\grg{\aaref@jnl{Gen.~Relat.~Gravit.}}               
\def\rpp{\aaref@jnl{Rep.~Prog.~Phys.}}          
\def\npa{\aaref@jnl{Nucl.~Phys.~A}}        
\def\lrr{\aaref@jnl{Living Rev.~Rel.}}                   
\def\jcap{\aaref@jnl{J.~Cosmology Astropart.~Phys.}}    
\def\rmp{\aaref@jnl{Rev.~Mod.~Phys.}}   
\def\epjc{\aaref@jnl{Eur.~Phys.~J.~C}}

\allowdisplaybreaks[1]

\begin{document}

\color{black}       

\title{A new parametrization of Hubble parameter in $f(Q)$ gravity}
\author{M. Koussour\orcidlink{0000-0002-4188-0572}}
\email{pr.mouhssine@gmail.com}
\affiliation{Quantum Physics and Magnetism Team, LPMC, Faculty of Science Ben
M'sik,\\
Casablanca Hassan II University,
Morocco.}

\author{S. K. J. Pacif\orcidlink{0000-0003-0951-414X}}
\email{shibesh.math@gmail.com}
\affiliation{Centre for Cosmology and Science Popularization (CCSP), SGT University, Delhi-NCR, Gurugram 122505, Haryana, India.}

\author{M. Bennai\orcidlink{0000-0002-7364-5171}}
\email{mdbennai@yahoo.fr}
\affiliation{Quantum Physics and Magnetism Team, LPMC, Faculty of Science Ben
M'sik,\\
Casablanca Hassan II University,
Morocco.} 
\affiliation{Lab of High Energy Physics, Modeling and Simulations, Faculty of
Science,\\
University Mohammed V-Agdal, Rabat, Morocco.}

\author{P.K. Sahoo\orcidlink{0000-0003-2130-8832}}
\email{pksahoo@hyderabad.bits-pilani.ac.in}
\affiliation{Department of Mathematics, Birla Institute of Technology and
Science-Pilani,\\ Hyderabad Campus, Hyderabad-500078, India.}
%
\date{\today}
\begin{abstract}
In this paper, we examine the accelerated expansion of the Universe at
late-time in the framework of $f\left( Q\right) $ gravity theory in which
the non-metricity scalar $Q$ describes the gravitational interaction. 
To this, we propose a new parametrization of the Hubble parameter
using a model-independent way and apply it to the Friedmann equations in the
FLRW Universe. Then we estimate the best fit values of the model parameters
by using the combined datasets of updated $H(z)$ consisting of $57$ points,
the Pantheon consisting of $1048$ points, and BAO datasets consisting of six
points with the Markov Chain Monte Carlo (MCMC) method. The evolution of
deceleration parameter indicates a transition from the deceleration to the
acceleration phase of the Universe. In addition, we investigate the behavior
of statefinder analysis and Om diagnostic parameter, Further, to discuss
other cosmological parameters, we consider a $f\left( Q\right) $\ model,
specifically, $f\left( Q\right) =Q+mQ^{n}$, where $m$ and $n$ are free
parameters. Finally, we find that the model supports the present
accelerating Universe, and the EoS parameter behaves like the quintessence
model.
\end{abstract}

\maketitle

\date{\today}

\section{Introduction}

\label{sec1}

Recent observational data of type Ia supernovae (SNIa) \cite{SN1,SN2}, Large
Scale Structure (LSS) \cite{LS1,LS2}, Wilkinson Microwave Anisotropy Probe
(WMAP) data \cite{WMAP1,WMAP2,WMAP9}, Cosmic Microwave Background (CMB) \cite%
{CMB1,CMB2}, and Baryonic Acoustic Oscillations (BAO) \cite{BAO1,BAO2}
confirm that the the expansion of the Universe has entered an acceleration
phase. Further, the same observational data exhibit that everything we see
around us represents only $5\%$ of the total content of the Universe, and
the remaining content, i.e. $95\%$ is in the form of other unknown
components of matter and energy called Dark Matter (DM) and Dark Energy
(DE). The results of these recent observations contradict the standard
Friedmann equations in General relativity (GR), which are part of the
applications of GR on a homogeneous and isotropic Universe on a large scale.
Thus, GR is not the ultimate theory of gravitation as we would have thought,
it may be a particular case of a more general theory.

Theoretically, to explain the current acceleration of the Universe, there
are two approaches in the literature. The first approach in the framework of
GR is to modify the content of the Universe by adding new components of
matter and energy such as DE of a nature unknown to this day, which has a
large negative pressure. The most famous cosmological constant ($\Lambda $)
that Einstein introduced into his field equations in GR is the most favored
candidate for DE as it fits very well observations. The idea is that the
cosmological constant was originated from the vacuum energy predicted by
quantum field theory \cite{Weinberg}. However, with this well motivated candidate of DE - the cosmological constant ($\Lambda$) suffers from some problems. The major one is it's constant
equation of state. In literature, there were several dynamical models of $%
\Lambda $ explored in order to resolve the cosmological constant problem
prior to the discovery of late-time accelerated expansion. With the
discovery of cosmic acceleration, the cosmological constant introduced into
the Einstein field equations (EFEs). Moreover, a slow roll scalar
field (potential dominated scalar field) reduces to the case of cosmological
constant. So, it is convenient to consider a scalar field, which has a
dynamical equation of state - the quintessence model \cite{quintessence}%
\textbf{.} Other dynamical (time-varying) models of DE have been proposed
such as phantom DE \cite{phantom}, k-essence \cite{kessence}, Chameleon \cite%
{Chameleon}, tachyon \cite{tachyon}, and Chaplygin gas \cite{Cgas1,Cgas2}.

The second approach to explain the current acceleration of the expansion of
the Universe is to modify the Einstein-Hilbert action in GR theory. The
fundamental concept in GR is the curvature imported from Riemannian geometry
which is described by the Ricci scalar $R$. The modified $f(R)$ gravity is a
simple modification of GR, replacing the Ricci scalar $R$ with some general
function of $R$\ \cite{fR}. Furthermore, there are other alternatives to GR
such as the teleparallel equivalent of GR (TEGR), in which the gravitational
interactions are described by the concept of torsion $T$. In GR, the
Levi-Civita connection is associated with curvature, but zero torsion, while
in teleparallelism, the Weitzenbock connection is associated with torsion,
but zero curvature \cite{fT}. In the same way, the $f\left( T\right) $
gravity is the simplest modification of TEGR. Recently, a new theory of
gravity has been proposed called the symmetric teleparallel equivalent of GR
(STEGR), in which the gravitational interactions are described by the
concept of non-metricity $Q$ with zero torsion and curvature \cite%
{Jimenez1,Jimenez2}. The non-metricity in Weyl geometry (generalization of
Riemannian geometry) represents the variation of a vector length in parallel
transport. In Weyl geometry, the covariant derivative of the metric tensor
is not equal to zero but is determined mathematically by the non-metricity
tensor i.e. $Q_{\gamma \mu \nu }=-\nabla _{\gamma }g_{\mu \nu }$ \cite{Xu}.
Also, the $f(Q) $ gravity is the simplest modification of STEGR. Many issues
have been discussed in the framework of $f(Q)$ gravity sufficiently enough
to motivate us to work under this new framework. Mandal et al. have examined
the energy conditions and cosmography in $f(Q)$ gravity \cite%
{Mandal1,Mandal2}, while Harko et al. investigated the coupling matter in
modified $Q$ gravity by presuming a power-law function \cite{Harko1}.
Dimakis et al. discussed quantum cosmology for a $f(Q)$ polynomial model 
\cite{Dimakis}, see also \cite{Koussour1,Koussour2}.

In this work, we consider a new simple parametrization of the Hubble
parameter to obtain the scenario of an accelerating Universe, with the study
of the most famous model of the function $f(Q)$ in the literature 
\cite{Mandal3} which contains a linear and a non-linear form of
non-metricity scalar, specifically $f\left( Q\right) =Q+mQ^{n}$,
where $m$ and $n$ are free parameters. The best fit
values of model parameters were obtained from the recent observational
datasets, which are mostly used in this topic. The Hubble datasets $H\left(
z\right) $ consisting of a list of $57$ measurements that were compiled from
the differential age method \cite{Sharov} and others, the Type Ia supernovae
sample called Pantheon datasets consisting $1048$ points covering the
redshift range $0.01<z<2.26$ \cite{Scolnic}, and BAO datasets consisting of
six points \cite{Blake}. Our analysis uses the combined of the $H(z)$,
Pantheon samples and BAO datasets to constrain the cosmological model.
Moreover, an MCMC (Markov Chain Monte Carlo) approach given by the emcee
library will be used to estimate parameters \cite{Mackey}.

The paper is organized as follows: In Sec. \ref{sec2}, we present an
overview of $f(Q)$ gravity theory in a flat FLRW Universe. In Sec. \ref{sec3}
we describe the cosmological parameters obtained from a new parametrization
of the Hubble parameter used to get the exact solutions of the field
equations. Further, we constraint the model parameters by using the combined
of the $H(z)$, Pantheon samples and BAO datasets. Next, we consider a $%
f\left( Q\right) $ cosmological model in Sec. \ref{sec4}. The behavior of
some cosmological parameters such as the energy density, pressure, and EoS
parameter are discussed in the same section. Finally, Sec. \ref{sec5} is
dedicated to conclusions.

\section{Overview of $f(Q)$ gravity theory}

\label{sec2}

In differential geometry, the metric tensor $g_{\mu \nu }$ is thought to be
a generalization of gravitational potentials. It is mainly used to determine
angles, distances and volumes, while the affine connection $\Upsilon {%
^{\gamma }}_{\mu \nu }$ has its main role in parallel transport and
covariant derivatives. In the case of Weyl geometry with the presence of the
non-metricity term, the general affine connection can be decomposed into the
following two independent components: the Christoffel symbol ${\Gamma
^{\gamma }}_{\mu \nu }$ and the disformation tensor ${L^{\gamma }}_{\mu \nu
} $ as follows \cite{Xu} 
\begin{equation}
\Upsilon {^{\gamma }}_{\mu \nu }={\Gamma ^{\gamma }}_{\mu \nu }+{L^{\gamma }}%
_{\mu \nu },  \label{WC}
\end{equation}%
where the Christoffel symbol is related to the metric tensor $g_{\mu \nu }$
by 
\begin{equation}
{\Gamma ^{\gamma }}_{\mu \nu }\equiv \frac{1}{2}g^{\gamma \sigma }\left(
\partial _{\mu }g_{\sigma \nu }+\partial _{\nu }g_{\sigma \mu }-\partial
_{\sigma }g_{\mu \nu }\right)
\end{equation}%
and the disformation tensor ${L^{\gamma }}_{\mu \nu }$ is derived from the
non-metricity tensor $Q_{\gamma \mu \nu }$ as 
\begin{equation}
{L^{\gamma }}_{\mu \nu }\equiv \frac{1}{2}g^{\gamma \sigma }\left( Q_{\nu
\mu \sigma }+Q_{\mu \nu \sigma }-Q_{\gamma \mu \nu }\right) .  \label{L}
\end{equation}

The non-metricity tensor $Q_{\gamma \mu \nu }$ is defined as the covariant
derivative of the metric tensor with regard to the Weyl connection $\Upsilon 
{^{\gamma }}_{\mu \nu }$, i.e. 
\begin{equation*}
Q_{\gamma \mu \nu }=-\nabla _{\gamma }g_{\mu \nu },
\end{equation*}%
and it can be obtained 
\begin{equation}
Q_{\gamma \mu \nu }=-\partial _{\gamma }g_{\mu \nu }+g_{\nu \sigma }\Upsilon 
{^{\sigma }}_{\mu \gamma }+g_{\sigma \mu }\Upsilon {^{\sigma }}_{\nu \gamma
}.
\end{equation}

The space-time in symmetric teleparallel gravity or the so-called $f(Q)$\
gravity is constructed using non-metricity and the symmetric teleparallelism
condition, where torsion and curvature both vanish i.e. $T_{\mu \nu
}^{\gamma }=0$ and $R_{\sigma \mu \nu }^{\rho }=0$. This
condition, as mentioned in Refs. \cite{Jim1, Jim2, Jim3}, allows to adopt a
coordinate system $\left\{ \xi ^{\mu }\right\} $\ so that the affine
connection $\Upsilon _{\mu \nu }^{\gamma }$\ disappears, which is the
so-called coincident gauge \cite{Jimenez1}. The affine connection then has
the following form in any other coordinate system $\left\{ x^{\mu }\right\} $%
,%
\begin{equation}
\Upsilon ^{\gamma }\,_{\mu \nu }\left( x^{\mu }\right) =\frac{\partial
x^{\gamma }}{\partial \xi ^{\beta }}\partial _{\mu }\partial _{\nu }\xi
^{\beta },
\end{equation}%
while in an arbitrary coordinate system, we have 
\begin{equation}
Q_{\gamma \mu \nu }=\partial _{\gamma }g_{\mu \nu }-2\Upsilon _{\gamma (\mu
}^{\lambda }g_{\nu )\lambda }.
\end{equation}

Further, the following relationship can be obtained between the curvature
tensors $R_{\sigma \mu \nu }^{\rho }$\ and $\mathring{R}_{\sigma \mu \nu
}^{\rho }$\ associated to the connection $\Upsilon _{\mu \nu }^{\gamma }$\
and $\Gamma _{\mu \nu }^{\gamma }$\ as,%
\begin{equation}
R_{\sigma \mu \nu }^{\rho }=\mathring{R}_{\sigma \mu \nu }^{\rho }+\mathring{%
\nabla}_{\mu }L_{\nu \sigma }^{\rho }-\mathring{\nabla}_{\nu }L_{\mu \sigma
}^{\rho }+L_{\mu \gamma }^{\rho }L_{\nu \sigma }^{\gamma }-L_{\nu \gamma
}^{\rho }L_{\mu \sigma }^{\gamma },
\end{equation}%
and%
\begin{equation}
R_{\sigma \nu }=\mathring{R}_{\sigma \nu }+\frac{1}{2}\mathring{\nabla}_{\nu
}Q_{\sigma }+\mathring{\nabla}_{\rho }L_{\nu \sigma }^{\rho }-\frac{1}{2}%
Q_{\gamma }L_{\nu \sigma }^{\gamma }-L_{\sigma \gamma }^{\rho }L_{\rho
\sigma }^{\gamma },
\end{equation}%
\begin{equation}
R=\mathring{R}+\mathring{\nabla}_{\gamma }Q^{\gamma }-\mathring{\nabla}%
_{\gamma }\tilde{Q}^{\gamma }-\frac{1}{4}Q_{\gamma }Q^{\gamma }+\frac{1}{2}%
Q_{\gamma }\tilde{Q}^{\gamma }-L_{\rho \nu \gamma }L^{\gamma \rho \nu }.
\end{equation}

Here $R_{\sigma \nu }$\ and $R$\ are, respectively, the Ricci tensor and the
Ricci scalar of the Weyl space. Since the affine connexion is zero in this
gauge, the curvature tensor is also zero which causes the overall geometry
of space-time to be flat. Thus, the covariant\ derivative $\nabla _{\gamma }$
reduces to the partial derivative $\partial _{\gamma }$ i.e. $Q_{\gamma \mu
\nu }=-\partial _{\gamma }g_{\mu \nu }$. It is clear from the previous
discussion that the Levi-Civita connection ${\Gamma ^{\gamma }}_{\mu \nu }$
can be written in terms of the disformation tensor ${L^{\gamma }}_{\mu \nu }$
as, 
\begin{equation}
{\Gamma ^{\gamma }}_{\mu \nu }=-{L^{\gamma }}_{\mu \nu }.
\end{equation}

The action for the gravitational interactions in symmetric teleparallel
geometry is given as \cite{Jimenez1,Jimenez2} 
\begin{equation}
S=\int \sqrt{-g}d^{4}x\left[ \frac{1}{2}f(Q)+\mathcal{L}_{m}\right] ,
\label{action}
\end{equation}%
where $f(Q)$ is an arbitrary function of non-metricity scalar $Q$, while $g$
is the determinant of the metric tensor $g_{\mu \nu }$ and $\mathcal{L}_{m}$
is the matter Lagrangian density. This above action, along with a flat and
symmetric connection constraint, determines the dynamics of the
gravitational field, imposing the vanishing of the total curvature of the
Weyl space-time $R_{\sigma \mu \nu }^{\rho }=0$. This constraint is imposed
by including a Lagrange multiplier in the gravitational action, see \cite%
{Jimenez1,Jim1}.

The trace of the non-metricity tensor $Q_{\gamma \mu \nu }$ can be written
as 
\begin{equation}
Q_{\gamma }={{Q_{\gamma }}^{\mu }}_{\mu }\,,\qquad \widetilde{Q}_{\gamma }={%
Q^{\mu }}_{\gamma \mu }\,.
\end{equation}

It is also useful to introduce the superpotential tensor (non-metricity
conjugate) given by 
\begin{equation}
4{P^{\gamma }}_{\mu \nu }=-{Q^{\gamma }}_{\mu \nu }+2Q{_{(\mu \;\;\nu
)}^{\;\;\;\gamma }}+Q^{\gamma }g_{\mu \nu }-\widetilde{Q}^{\gamma }g_{\mu
\nu }-\delta _{\;(\mu }^{\gamma }Q_{\nu )}\,,
\end{equation}%
where the trace of the non-metricity tensor can be obtained as 
\begin{equation}
Q=-Q_{\gamma \mu \nu }P^{\gamma \mu \nu }\,.
\end{equation}

The symmetric teleparallel gravity field equations are obtained by varying
the action $S$ with respect to the metric tensor $g_{\mu \nu }$, 
\begin{multline}
\frac{2}{\sqrt{-g}}\nabla _{\gamma }(\sqrt{-g}f_{Q}P^{\gamma }{}_{\mu \nu })+%
\frac{1}{2}fg_{\mu \nu } \\
+f_{Q}(P_{\nu \rho \sigma }Q_{\mu }{}^{\rho \sigma }-2P_{\rho \sigma \mu
}Q^{\rho \sigma }{}_{\nu })=-T_{\mu \nu }.  \label{F}
\end{multline}%
where the energy-momentum tensor is given by 
\begin{equation}
T_{\mu \nu }=-\frac{2}{\sqrt{-g}}\frac{\delta (\sqrt{-g}\mathcal{L}_{m})}{%
\delta g^{\mu \nu }},
\end{equation}%
$f_{Q}={df}/{dQ}$ and $\nabla _{\mu }$ denotes the covariant derivative. By
varying the action with respect to the affine connection $\Upsilon _{\mu \nu
}^{\gamma }$, we find 
\begin{equation}
\nabla ^{\mu }\nabla ^{\nu }\left( \sqrt{-g}\,f_{Q}\,P^{\gamma }\;_{\mu \nu
}\right) =0.  \label{conn}
\end{equation}

It is important to note that Eq. (\ref{conn}) is only valid in the absence
of hypermomentum. Also, the Bianchi identity implies that this equation is
automatically satisfied once the metric equations of motion are satisfied 
\cite{Jim4}.

The cosmological principle states that our Universe is homogeneous and
isotropic in the large scale. The mathematical description of a homogeneous
and isotropic Universe is given by the flat
Friedmann-Lemaitre-Robertson-Walker (FLRW) metric. So for this metric,
according to \cite{Jimenez2}, coincident gauges are consistent with the
Cartesian coordinate system, which indicates that in the Cartesian
coordinate system, selecting $\Upsilon _{\mu \nu }^{\gamma }=0$\ is a
solution of $f(Q)$\ theory. The FLRW metric is regarded as, 
\begin{equation}
ds^{2}=-dt^{2}+a^{2}(t)\left[ dx^{2}+dy^{2}+dz^{2}\right] ,  \label{FLRW}
\end{equation}%
where $a(t)$ is the scale factor that measures the size of the expanding
Universe. From now on, and unless otherwise mentioned, we will fix the
coincident gauge such that the connection is trivial and the metric is just
a fundamental variable.

The non-metricity scalar corresponding to the spatially flat FLRW line
element is obtained as 
\begin{equation}
Q=6H^{2},
\end{equation}%
where $H$ is the Hubble parameter which measures the rate of expansion of
the Universe.

To obtain the modified Friedmann equations that govern the Universe when it
is described by the spatially flat FLRW metric, we use the stress-energy
momentum tensor most commonly used in cosmology, i.e. the stress-energy
momentum tensor of perfect fluid given by 
\begin{equation}
T_{\mu \nu }=(p+\rho )u_{\mu }u_{\nu }+pg_{\mu \nu },  \label{PF}
\end{equation}%
where $p$ and $\rho $ represent the isotropic pressure and the energy
density of the perfect fluid, respectively. Here, $u^{\mu }=\left(
1,0,0,0\right) $ are components of the four velocities of the perfect fluid.

In view of Eq. (\ref{PF}) for the spatially flat FLRW metric, the symmetric
teleparallel gravity field equations (\ref{F}) lead to 
\begin{equation}
3H^{2}=\frac{1}{2f_{Q}}\left( -\rho +\frac{f}{2}\right) ,  \label{F1}
\end{equation}%
\begin{equation}
\dot{H}+3H^{2}+\frac{\dot{f}_{Q}}{f_{Q}}H=\frac{1}{2f_{Q}}\left( p+\frac{f}{2%
}\right) ,  \label{F2}
\end{equation}%
where the dot $(\overset{.}{})$ denotes the derivative with respect to the
cosmic time $t$.

Now, by eliminating the term $3H^{2}$\ from the previous two equations, we
get the following evolution equation for $H$, 
\begin{equation}
\dot{H}+\frac{\dot{f}_{Q}}{f_{Q}}H=\frac{1}{2f_{Q}}\left( p+\rho \right) .
\label{evo}
\end{equation}

Using the above equations, Eqs. (\ref{F1}) and (\ref{F2}), we obtain the
expressions of the energy density $\rho $ and the isotropic pressure $p$,
respectively as 
\begin{equation}
\rho =\frac{f}{2}-6H^{2}f_{Q},  \label{F22}
\end{equation}%
\begin{equation}
p=\left( \dot{H}+3H^{2}+\frac{\dot{f_{Q}}}{f_{Q}}H\right) 2f_{Q}-\frac{f}{2}.
\label{F33}
\end{equation}

Again, using Eqs. (\ref{F2}) and (\ref{evo}) we can rewrite the cosmological
equations similar to the standard Friedmann equations in GR, by adding the
concept of an effective energy density $\rho _{eff}$ and an effective
isotropic pressure $p_{eff}$ as, 
\begin{equation}
3H^{2}=\rho _{eff}\,=-\frac{1}{2f_{Q}}\left( \rho -\frac{f}{2}\right) ,
\label{eff1}
\end{equation}%
\begin{equation}
2\dot{H}+3H^{2}=-p_{eff}\,=-\frac{2\dot{f_{Q}}}{f_{Q}}H+\frac{1}{2f_{Q}}%
\left( \rho +2p+\frac{f}{2}\right) .  \label{eff2}
\end{equation}

Furthermore, the gravitational action (\ref{action}) is reduced to the
standard Hilbert-Einstein form in the limiting case $f\left( Q\right)
=-Q=-6H^{2}$. For this choice, we regain the so-called STEGR \cite{Lazkoz},
and Eqs. (\ref{eff1}) and (\ref{eff2}) reduce to the standard Friedmann
equations of GR, $3H^{2}=\rho $, and $2\dot{H}+3H^{2}=-p$, respectively.

\section{New parametrization of the Hubble parameter}

\label{sec3}

Generally, the above system of field equations consists of only two
independent equations with four unknowns, namely $\rho $, $p$, $f\left(
Q\right) $, $H$. In order to study the time evolution of the Universe and
some cosmological parameters. Further, from a mathematical point of view, to
solve the system completely we need additional constraints. In literature,
there are many physical motivations for choosing these constraints, the most
famous of which is the model-independent method to study the dynamics of
dark energy models \cite{Shafieloo}. The principle of this approach is to
consider a parametrization of any cosmological parameters such as the Hubble
parameter, deceleration parameter, and equation of state (EoS)\ parameter.
Hence, the necessary supplementary equation has been provided. Sahni et al. 
\cite{Sahni} discussed the statefinder diagnostic by assuming a
model-independent parametrization of the Hubble parameter $H\left( z\right) $%
. Recently, the same parametrization of $H\left( z\right) $ in modified $%
f(Q) $ gravity was discussed by Mandal et al. \cite{Mandal3}. Further,
several parametrizations have been investigated for the EoS parameter $%
\omega \left( z\right) $ such as CPL (Chevallier-Polarski-Linder), BA
(Barboza-Alcaniz), LC (Low Correlation) \cite{Escamilla-Rivera}, and the
deceleration parameter $q\left( z\right) $, see \cite{Banerjee,Cunha}. In
addition to the parametrization of the deceleration parameter, there are
several other schemes of parametrization of other cosmological parameters.
These schemes have been extensively addressed in the literature to describe
issues with cosmological investigations e.g. initial singularity problem,
problem all-time decelerating expansion problem, horizon problem, Hubble
tension etc. For a detailed review of the various schemes of cosmological
parameterization, one may follow \cite{PACIF1,PACIF2}. These investigations,
motivated us to work with a new parametrization of the Hubble parameter in
the form

\begin{equation}
H\left( z\right) =H_{0}\left[ \left( 1-\alpha \right) +\left( 1+z\right)
\left( \alpha +\beta z\right) \right] ^{\frac{1}{2}},  \label{Hz}
\end{equation}%
where $\alpha $ and $\beta $ are model parameters and can be measured using
observational data, $H_{0}$ is the Hubble value at $z=0$.

The deceleration parameter is one of the cosmological parameters that play
an important role in describing the state of expansion of our Universe. The
cosmological models of the evolution of the Universe transit from the early
deceleration phase ($q>0$) to the present accelerated phase ($q<0$) with
certain values of the transition redshift $z_{t}$. Further, the
observational data used in this paper showed that our current Universe
entered an accelerating phase with a deceleration parameter ranging between $%
-1\leq q\leq 0$. The deceleration parameter is defined in terms of the
Hubble parameter as 
\begin{equation}
q\left( z\right) =-1-\frac{\overset{.}{H}}{H^{2}}.  \label{qz}
\end{equation}

In addition, using the relation between the redshift and the scale factor of
the Universe $a\left( t\right) =\left( z+1\right) ^{-1}$, we can define the
relation between the cosmic time and redshift as%
\begin{equation}
\frac{d}{dt}=\frac{dz}{dt}\frac{d}{dz}=-\left( z+1\right) H\left( z\right) 
\frac{d}{dz}.
\end{equation}

Using the above equation, the Hubble parameter can be written in the form%
\begin{equation}
\overset{.}{H}=-\left( z+1\right) H\left( z\right) \frac{dH}{dz}.  \label{Hd}
\end{equation}

Now, using Eqs. (\ref{Hz}) and (\ref{qz}) and with the help of Eq. (\ref{Hd}%
), the deceleration parameter $q\left( z\right) $ according to our model is
given by, 
\begin{equation}
q\left( z\right) =\frac{\alpha +\beta +\alpha (-z)+\beta z-2}{2\left( \beta
z^{2}+z(\alpha +\beta )+1\right) }.
\end{equation}

To study the behavior of the cosmological parameters above, the next step
will be to find the best fit values of the model parameters $\alpha $ and $%
\beta $ using the combination of the $H(z)$, Pantheon samples and BAO
datasets.

\subsection{Observational constraints}

In the previous sections, we have briefly described the $f(Q)$ gravity and
solved the field equation with a new parametrization of Hubble parameter.
The considered form of $H(z)$ contains two model parameters $\alpha $ \& $%
\beta $, which have been constrained through some observational data for
further analysis. We have used some external datasets, such as observational
Hubble datasets of recently compiled $57$ data points, Pantheon compilation
of SNe\textit{Ia} data with $1048$ data points, and also the Baryonic
Acoustic Oscillation datasets with six data points, to obtain the best fit
values for these model parameters in order to validate our technique. In
order to limit the model parameters, we have first used the scipy
optimization technique from Python library to determine the global minima
for the considered Hubble function in equation (\ref{Hz}). It is apparent
that the parameters' diagonal covariance matrix entries have significant
variances. The aforesaid estimations were then taken into account as means
and a Gaussian prior with a fixed $\sigma =1.0$ as the dispersion was
utilised for the numerical analysis using Python's emcee package. Given
this, we examined the parameter space encircling the local minima (or
estimates). Below, a more in-depth analysis of the technique used with three
datasets is provided. The results are shown in the contour plots
(two-dimensional) with $1-\sigma $ and $2-\sigma $ errors.

\subsubsection{Hz datasets}

As a function of redshift, the Hubble parameter may be written as $H(z)$ $=-%
\frac{1}{1+z}\frac{dz}{dt}$, where $dz$ is determined from spectroscopic
surveys. In contrast, determining $dt$ yields the Hubble parameter's
model-independent value. The value of the $H(z)$ at a certain redshift is
frequently estimated using two methods. The differential age methodology is
one, while the extraction of H(z) from the line-of-sight BAO data is another 
\cite{h1}-\cite{h19}. The reference \cite{Sharov} provides a quick summary
of a list of revised datasets of $57$ points out of which $31$ points from
the differential age technique and the remaining $26$ points evaluated using
BAO and other methods in the redshift range of $0.07\leqslant z\leqslant
2.42 $. Furthermore, we have used $H_{0}=69$ Km/s/Mpc for our analysis. The
maximum likelihood analysis's counterpart, the chi-square function, is used
to determine the model parameters' average values and is given by,

\begin{equation}
\chi _{H}^{2}(\alpha ,\beta )=\sum\limits_{i=1}^{57}\frac{%
[H_{th}(z_{i},\alpha ,\beta )-H_{obs}(z_{i})]^{2}}{\sigma _{H(z_{i})}^{2}},
\label{chihz}
\end{equation}%
where $H_{obs}$ denotes the observed value of the Hubble parameter and $%
H_{th}$ denotes its theorised value. The symbol $\sigma _{H(z_{i})}$ denotes
the standard error in the observed value of the Hubble parameter $H(z)$. The
following Table $1$ described the $57$ points of the Hubble parameter values 
$H(z)$ with corresponding errors $\sigma _{H}$ from differential age ($31$
points), BAO and other ($26$ points), methods.

\begin{table*}[htbp]
\centering
\begin{tabular}{|c|c|c|c|c|c|c|c|}
\hline
\multicolumn{8}{|c|}{Table-1: 57 points of Hubble ($H(z)$) datasets} \\ 
\hline
\multicolumn{8}{|c|}{31 points of $H(z)$ datasets by DA method} \\ \hline
$z$ & $H(z)$ & $\sigma _{H}$ & Ref. & $z$ & $H(z)$ & $\sigma _{H}$ & Ref. \\ 
\hline
$0.070$ & $69$ & $19.6$ & \cite{h1} & $0.4783$ & $80$ & $99$ & \cite{h5} \\ 
\hline
$0.90$ & $69$ & $12$ & \cite{h2} & $0.480$ & $97$ & $62$ & \cite{h1} \\ 
\hline
$0.120$ & $68.6$ & $26.2$ & \cite{h1} & $0.593$ & $104$ & $13$ & \cite{h3}
\\ \hline
$0.170$ & $83$ & $8$ & \cite{h2} & $0.6797$ & $92$ & $8$ & \cite{h3} \\ 
\hline
$0.1791$ & $75$ & $4$ & \cite{h3} & $0.7812$ & $105$ & $12$ & \cite{h3} \\ 
\hline
$0.1993$ & $75$ & $5$ & \cite{h3} & $0.8754$ & $125$ & $17$ & \cite{h3} \\ 
\hline
$0.200$ & $72.9$ & $29.6$ & \cite{h4} & $0.880$ & $90$ & $40$ & \cite{h1} \\ 
\hline
$0.270$ & $77$ & $14$ & \cite{h2} & $0.900$ & $117$ & $23$ & \cite{h2} \\ 
\hline
$0.280$ & $88.8$ & $36.6$ & \cite{h4} & $1.037$ & $154$ & $20$ & \cite{h3}
\\ \hline
$0.3519$ & $83$ & $14$ & \cite{h3} & $1.300$ & $168$ & $17$ & \cite{h2} \\ 
\hline
$0.3802$ & $83$ & $13.5$ & \cite{h5} & $1.363$ & $160$ & $33.6$ & \cite{h7}
\\ \hline
$0.400$ & $95$ & $17$ & \cite{h2} & $1.430$ & $177$ & $18$ & \cite{h2} \\ 
\hline
$0.4004$ & $77$ & $10.2$ & \cite{h5} & $1.530$ & $140$ & $14$ & \cite{h2} \\ 
\hline
$0.4247$ & $87.1$ & $11.2$ & \cite{h5} & $1.750$ & $202$ & $40$ & \cite{h2}
\\ \hline
$0.4497$ & $92.8$ & $12.9$ & \cite{h5} & $1.965$ & $186.5$ & $50.4$ & \cite%
{h7} \\ \hline
$0.470$ & $89$ & $34$ & \cite{h6} &  &  &  &  \\ \hline
\multicolumn{8}{|c|}{26 points of $H(z)$ datasets from BAO \& other method}
\\ \hline
$z$ & $H(z)$ & $\sigma _{H}$ & Ref. & $z$ & $H(z)$ & $\sigma _{H}$ & Ref. \\ 
\hline
$0.24$ & $79.69$ & $2.99$ & \cite{h8} & $0.52$ & $94.35$ & $2.64$ & \cite%
{h10} \\ \hline
$0.30$ & $81.7$ & $6.22$ & \cite{h9} & $0.56$ & $93.34$ & $2.3$ & \cite{h10}
\\ \hline
$0.31$ & $78.18$ & $4.74$ & \cite{h10} & $0.57$ & $87.6$ & $7.8$ & \cite{h14}
\\ \hline
$0.34$ & $83.8$ & $3.66$ & \cite{h8} & $0.57$ & $96.8$ & $3.4$ & \cite{h15}
\\ \hline
$0.35$ & $82.7$ & $9.1$ & \cite{h11} & $0.59$ & $98.48$ & $3.18$ & \cite{h10}
\\ \hline
$0.36$ & $79.94$ & $3.38$ & \cite{h10} & $0.60$ & $87.9$ & $6.1$ & \cite{h13}
\\ \hline
$0.38$ & $81.5$ & $1.9$ & \cite{h12} & $0.61$ & $97.3$ & $2.1$ & \cite{h12}
\\ \hline
$0.40$ & $82.04$ & $2.03$ & \cite{h10} & $0.64$ & $98.82$ & $2.98$ & \cite%
{h10} \\ \hline
$0.43$ & $86.45$ & $3.97$ & \cite{h8} & $0.73$ & $97.3$ & $7.0$ & \cite{h13}
\\ \hline
$0.44$ & $82.6$ & $7.8$ & \cite{h13} & $2.30$ & $224$ & $8.6$ & \cite{h16}
\\ \hline
$0.44$ & $84.81$ & $1.83$ & \cite{h10} & $2.33$ & $224$ & $8$ & \cite{h17}
\\ \hline
$0.48$ & $87.79$ & $2.03$ & \cite{h10} & $2.34$ & $222$ & $8.5$ & \cite{h18}
\\ \hline
$0.51$ & $90.4$ & $1.9$ & \cite{h12} & $2.36$ & $226$ & $9.3$ & \cite{h19}
\\ \hline
\end{tabular}%
\end{table*}

\subsubsection{Pantheon datasets}

The recent supernovae type Ia dataset, with $1048$ data points, the \textit{%
pantheon} sample \cite{DM/2018} is used here for stronger constraints of the
model parameters. The sample is the spectroscopically verified SNe Ia data
points, which spans the redshift range $0.01<z<2.26$. These informational
points provide an estimate of the distance moduli $\mu _{i}=\mu _{i}^{obs}$
in the redshift range $0<z_{i}\leq 1.41$. To determine which value of the
distance modulus fits our model parameters of the generated model best, we
compare the theoretical $\mu _{i}^{th}$ value and observed $\mu _{i}^{obs}$
value. The distance moduli are the logarithms $\mu _{i}^{th}=\mu
(D_{L})=m-M=5\log _{10}(D_{L})+\mu _{0}$, where $m$ and $M$ stand for
apparent and absolute magnitudes, respectively, and $\mu _{0}=5\log \left(
H_{0}^{-1}/Mpc\right) +25$ is the marginalised nuisance parameter. The
luminosity distance is seen as being,

\begin{eqnarray*}
D_{l}(z) &=&\frac{c(1+z)}{H_{0}}S_{k}\left( H_{0}\int_{0}^{z}\frac{1}{%
H(z^{\ast })}dz^{\ast }\right) , \\
\text{where }S_{k}(x) &=&\left\{ 
\begin{array}{c}
\sinh (x\sqrt{\Omega _{k}})/\Omega _{k}\text{, }\Omega _{k}>0 \\ 
x\text{, \ \ \ \ \ \ \ \ \ \ \ \ \ \ \ \ \ \ \ \ \ \ \ }\Omega _{k}=0 \\ 
\sin x\sqrt{\left\vert \Omega _{k}\right\vert })/\left\vert \Omega
_{k}\right\vert \text{, }\Omega _{k}<0%
\end{array}%
\right.
\end{eqnarray*}%
Here, $\Omega _{k}=0$ (flat space-time). To quantify the discrepancy between
the SN \textit{Ia} observational data and our model's predictions, we
calculated distance $D_{l}(z)$ and the chi square function $\chi _{SN}^{2}$.
For the Pantheon datasets, the $\chi _{SN}^{2}$ function is assumed to be,

\begin{equation}
\chi _{SN}^{2}(\mu _{0},\alpha ,\beta )=\sum\limits_{i=1}^{1048}\frac{[\mu
^{th}(\mu _{0},z_{i},\alpha ,\beta )-\mu ^{obs}(z_{i})]^{2}}{\sigma _{\mu
(z_{i})}^{2}},  \label{chisn}
\end{equation}%
$\sigma _{\mu (z_{i})}^{2}$ being the standard error in the observed value.

\subsubsection{BAO datasets}

The early Universe is the subject of the analysis of baryonic acoustic
oscillations (BAO). Thompson scattering establishes a strong bond between
baryons and photons in the early cosmos, causing them to act as a single
fluid that defies gravity and oscillates instead due to the intense pressure
of photons. The characteristic scale of BAO can be found by looking at the
sound horizon $r_{s}$ at the photon decoupling epoch $z_{\ast }$, and is
provided by the relation,

\begin{equation*}
r_{s}(z_{\ast })=\frac{c}{\sqrt{3}}\int_{0}^{\frac{1}{1+z_{\ast }}}\frac{da}{%
a^{2}H(a)\sqrt{1+(3\Omega _{0b}/4\Omega _{0\gamma })a}},
\end{equation*}%
where the quantities $\Omega _{0b}$ and $\Omega _{0\gamma }$ stand for the
baryon and photon densities, respectively, at the present time.
Additionally, the sound horizon scale can be used to derive the functions of
redshift $z$ for the angular diameter distance $D_{A}$ and $H$ (Hubble
expansion rate). Here, $d_{A}(z)$ is the co-moving angular diameter distance
related to $H(z)$ and is related as $d_{A}(z)=\int_{0}^{z}\frac{dz^{\prime }%
}{H(z^{\prime })}$. It is used to calculate the measured angular separation
of the BAO ($\triangle \theta $), where $\triangle \theta =\frac{r_{s}}{%
d_{A}(z)}$ in the 2 point correlation function of the galaxy distribution on
the sky and the measured redshift separation of the BAO ($\triangle z$),
where $\triangle z=H(z)r_{s}$. In this work, BAO datasets of $d_{A}(z_{\ast
})/D_{V}(z_{BAO})$ are taken from the references \cite%
{BAO1,BAO2,BAO3,BAO4,BAO5,BAO6} and the photon decoupling redshift ($z_{\ast
}$) is considered, $z_{\ast }\approx 1091$. Also, the term $D_{V}(z)$
defined by $D_{V}(z)=\left( d_{A}(z)^{2}z/H(z)\right) ^{1/3}$ is the
dilation scale. For this analysis, we have used the data as considered in 
\cite{BAO6}, which is described in the Table $2$.

\begin{table*}[htbp]
\centering
\begin{tabular}{|c|c|c|c|c|c|c|}
\hline
\multicolumn{7}{|c|}{Table-2: The values of $d_{A}(z_{\ast })/D_{V}(z_{BAO})$
for different values of $z_{BAO}$} \\ \hline
$z_{BAO}$ & $0.106$ & $0.2$ & $0.35$ & $0.44$ & $0.6$ & $0.73$ \\ \hline
$\frac{d_{A}(z_{\ast })}{D_{V}(z_{BAO})}$ & $30.95\pm 1.46$ & $17.55\pm 0.60$
& $10.11\pm 0.37$ & $8.44\pm 0.67$ & $6.69\pm 0.33$ & $5.45\pm 0.31$ \\ 
\hline
\end{tabular}%
\end{table*}

\qquad Now, the chi-square function for the BAO is given by \cite{BAO6}, 
\begin{equation}
\chi _{BAO}^{2}=X^{T}C^{-1}X\,,  \label{chibao}
\end{equation}
where

\begin{widetext}
\begin{equation*}
X=\left( 
\begin{array}{c}
\frac{d_{A}(z_{\star })}{D_{V}(0.106)}-30.95 \\ 
\frac{d_{A}(z_{\star })}{D_{V}(0.2)}-17.55 \\ 
\frac{d_{A}(z_{\star })}{D_{V}(0.35)}-10.11 \\ 
\frac{d_{A}(z_{\star })}{D_{V}(0.44)}-8.44 \\ 
\frac{d_{A}(z_{\star })}{D_{V}(0.6)}-6.69 \\ 
\frac{d_{A}(z_{\star })}{D_{V}(0.73)}-5.45%
\end{array}%
\right) \,, 
\end{equation*}

and the inverse covariance matrix $C^{-1}$, which is defined in the ref. 
\cite{BAO6} is given by,

\begin{equation*}
C^{-1}=\left( 
\begin{array}{cccccc}
0.48435 & -0.101383 & -0.164945 & -0.0305703 & -0.097874 & -0.106738 \\ 
-0.101383 & 3.2882 & -2.45497 & -0.0787898 & -0.252254 & -0.2751 \\ 
-0.164945 & -2.454987 & 9.55916 & -0.128187 & -0.410404 & -0.447574 \\ 
-0.0305703 & -0.0787898 & -0.128187 & 2.78728 & -2.75632 & 1.16437 \\ 
-0.097874 & -0.252254 & -0.410404 & -2.75632 & 14.9245 & -7.32441 \\ 
-0.106738 & -0.2751 & -0.447574 & 1.16437 & -7.32441 & 14.5022%
\end{array}%
\right) \, 
\end{equation*}
\end{widetext}

\begin{figure*}[htbp]
\centering
\includegraphics[scale=1.2]{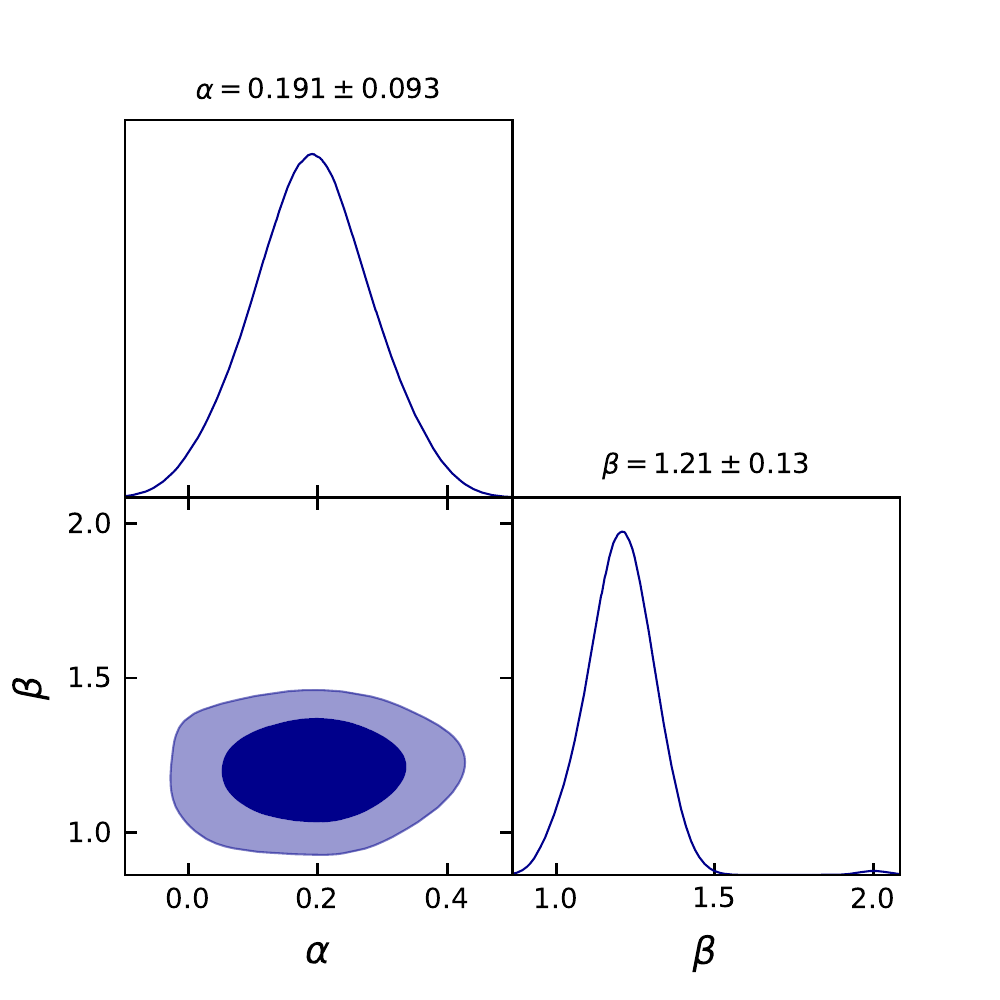}
\caption{The $1-\protect\sigma $ and $2-\protect\sigma $ likelihood contours
for the model parameters using $H(z)$+Pantheon+BAO datasets.}
\label{contour}
\end{figure*}

With the above set up, we have found the best fit values of the model
parameters for the combined Hubble, Pantheon and BAO datasets as $\alpha
=0.191_{-0.093}^{+0.093}$, $\beta =1.21_{-0.13}^{+0.13}$. The result is
shown in Fig. \ref{contour} as a two dimensional contour plots with $%
1-\sigma $ and $2-\sigma $ errors.

\begin{figure*}[htbp]
\centering
\includegraphics[scale=0.65]{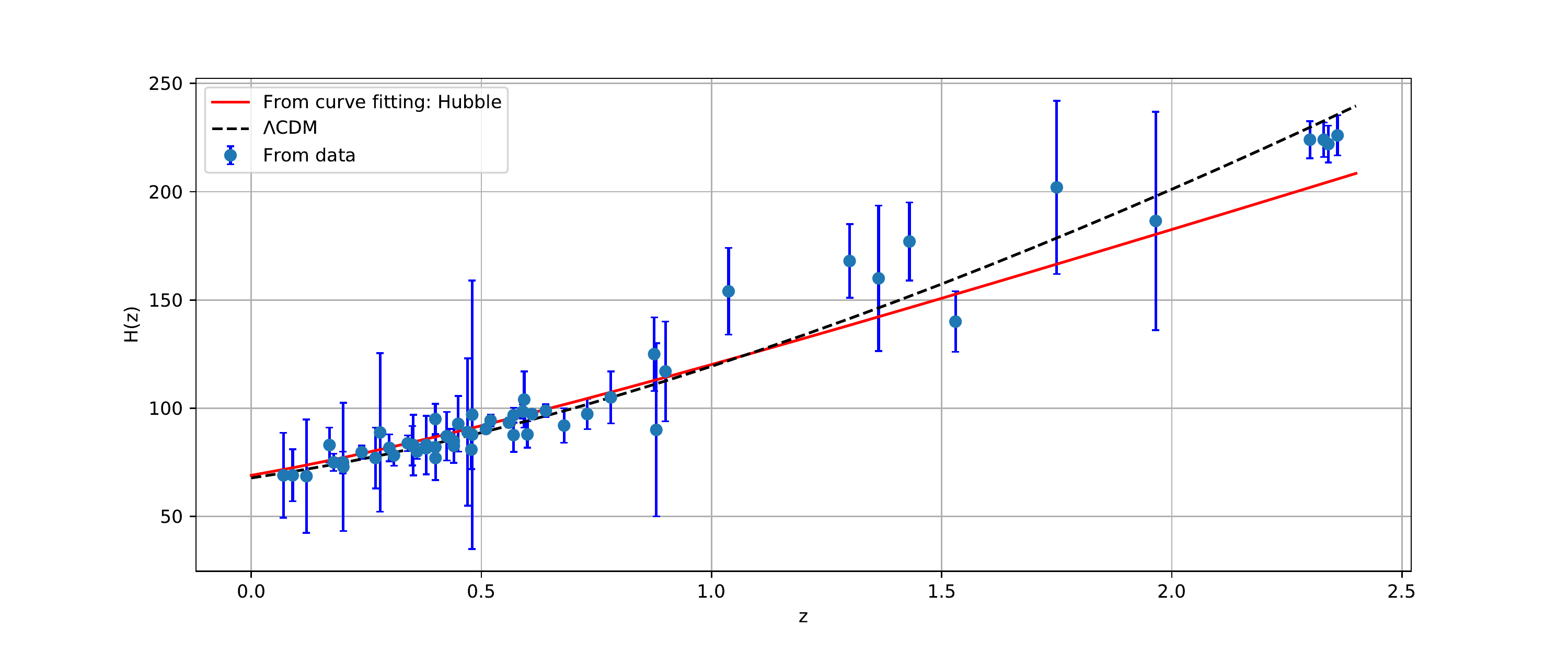}
\caption{The figure shows the error bar plot of the considered $57$ points
of Hubble datasets together with the fitting of Hubble function $H(z)$ vs.
redshift $z$ for our obtained model (red line) compared with that of
standard $\Lambda $CDM model (black dashed line).}
\label{Error-Hz}
\end{figure*}

\begin{figure*}[htbp]
\centering
\includegraphics[scale=0.65]{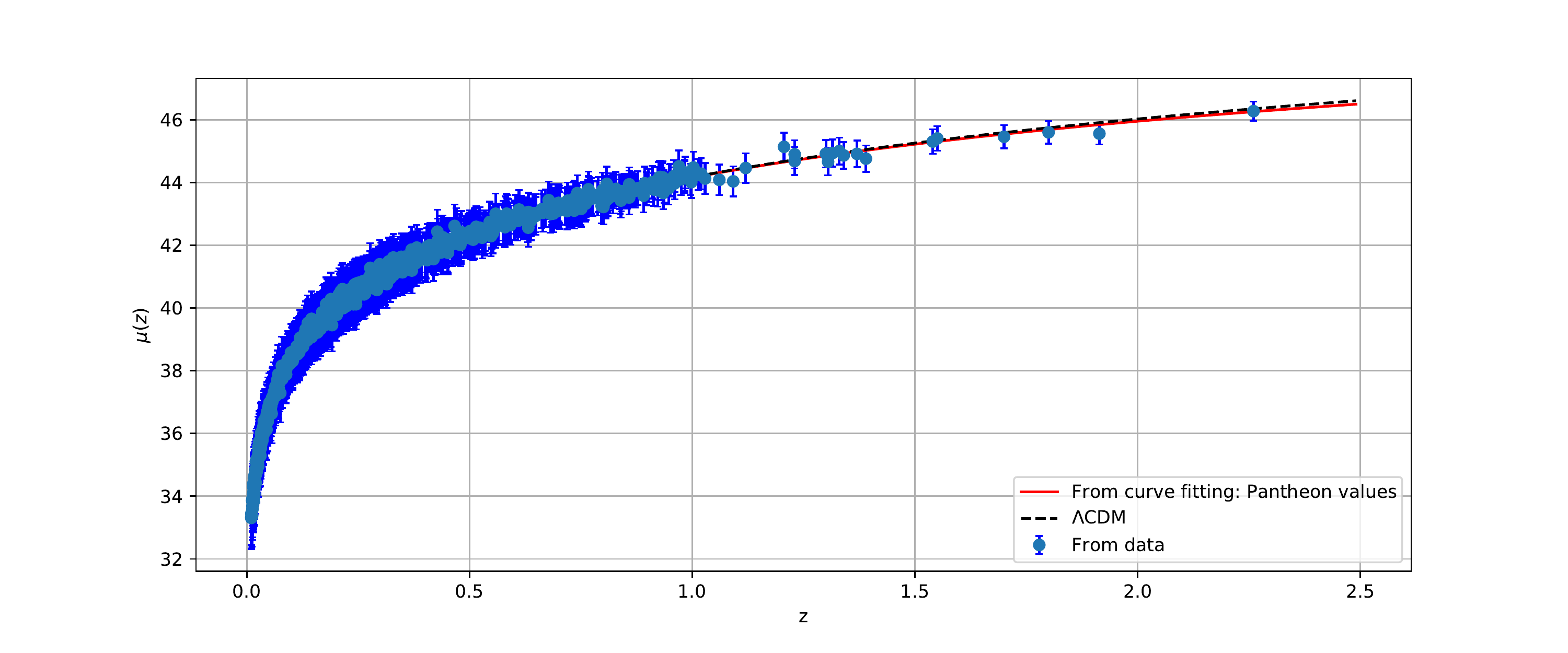}
\caption{The figure shows the error bar plot of the considered $1048$ points
of the Pantheon compilation SNe \textit{Ia }datasets together with the
fitting of function $\protect\mu (z)$ vs. redshift $z$ for our obtained
model (red line) compared with that of standard $\Lambda $CDM model (black
dashed line).}
\label{Error-Muz}
\end{figure*}

Additionally, we observed our derived model has nice fit to the
aforementioned Hubble and Pantheon datasets. The error bars for the
considered datasets and the $\Lambda $CDM model (with $\Omega _{\Lambda
0}=0.7$ and $\Omega _{m0}=0.3$) are also plotted along with our model for
comparison. This is displayed in Fig. \ref{Error-Hz} and \ref{Error-Muz}
respectively,

\subsection{Evolution of the $q(z)$ and phase transition}

\begin{figure}[]
\centerline{\includegraphics[scale=0.7]{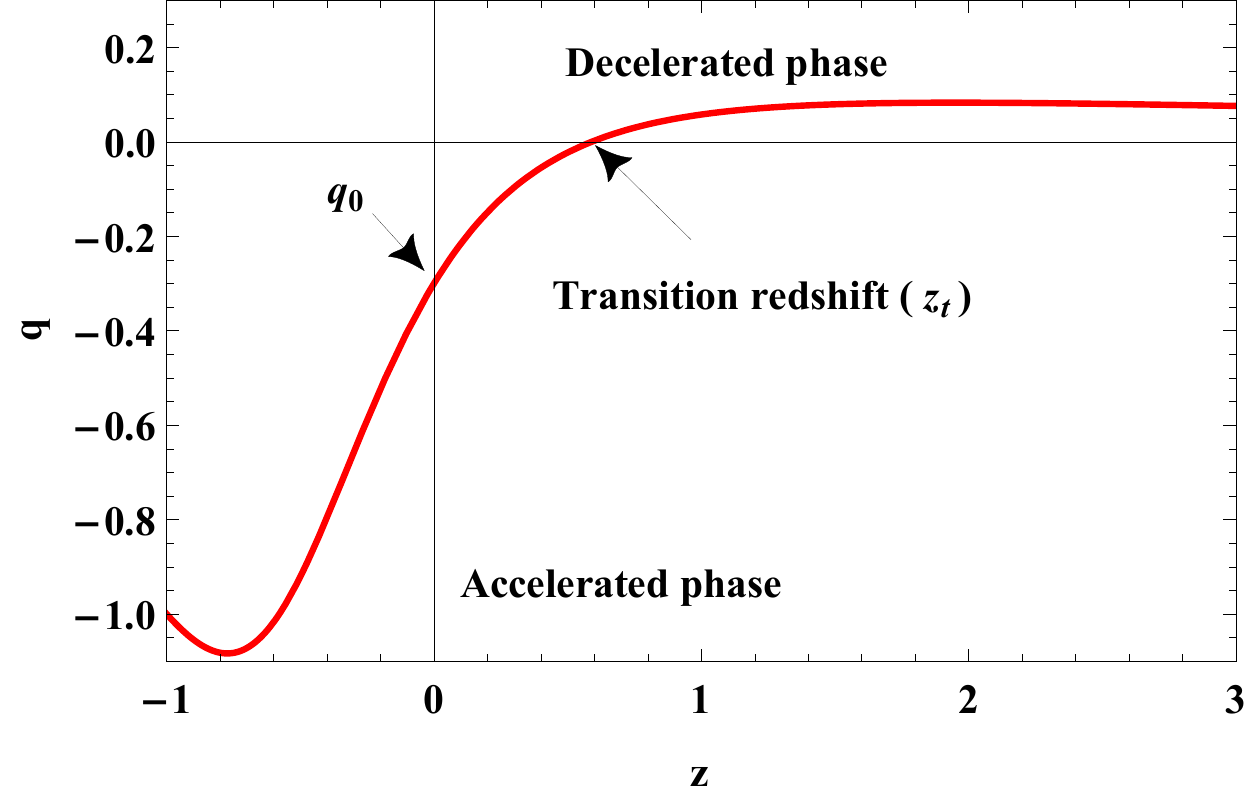}}
\caption{The graphical behavior of the deceleration parameter with the
constraint values from $H(z)$+Pantheon+BAO datasets.}
\label{fig_q}
\end{figure}

The evolution of the deceleration parameter corresponding to the constrained
values of the model parameters is shown in Fig. \ref{fig_q}. It is clear
from this figure that the cosmological model contains a transition from the
phase of deceleration to acceleration. The transition redshift corresponding
to the values of the model parameters constrained by the combined Hubble,
Pantheon and BAO datasets is $z_{t}=0.6857$. Moreover, the present value of
the deceleration parameter is $q_{0}=-0.3092$. Now, we are now fully
equipped with all theoretical formulas as well as numerical values of the
model parameters and can discuss the physical dynamics of the model. So, the
next section is dedicated to the physical dynamics of the other important
cosmological parameters.

\subsection{Statefinder analysis}

As mentioned above, the deceleration parameter plays a key role in knowing
the nature of the expansion of the Universe. But as more and more models are
presented for DE, the deceleration parameter no longer tells us enough about
the nature of the cosmological model, because DE models have the same
current value of this parameter. For this reason, it has become necessary to
propose new parameters to distinguish between DE models. Sahni et al.
proposed a new geometrical diagnostic parameters which are dimensionless and
known as statefinder parameters $\left( r,s\right) $ \cite{Sahni,Alam}. The
statefinder parameters are defined as%
\begin{equation}
r=\frac{\overset{...}{a}}{aH^{3}},
\end{equation}%
\begin{equation}
s=\frac{\left( r-1\right) }{3\left( q-\frac{1}{2}\right) }.
\end{equation}

The parameter $r$ can be rewritten as%
\begin{equation}
r=2q^{2}+q-\frac{\overset{.}{q}}{H}.
\end{equation}

For different values of the statefinder pair $\left( r,s\right) $, the
various DE models known in the literature can be represented as follows

\begin{itemize}
\item $\Lambda $CDM model corresponds to ($r=1,s=0$),

\item Chaplygin Gas (CG) model corresponds to ($r>1,s<0$),

\item Quintessence model corresponds to ($r<1,s>0$),
\end{itemize}

For our cosmological model, the statefinder pair can be obtained as%
\begin{equation}
r\left( z\right) =\frac{1-\alpha }{\beta z^{2}+z(\alpha +\beta )+1},
\end{equation}%
\begin{equation}
s\left( z\right) =\frac{2(z+1)(\alpha +\beta z)}{3\beta \left(
z^{2}-1\right) +\alpha (6z-3)+9}.
\end{equation}

\begin{figure}[]
\centerline{\includegraphics[scale=0.7]{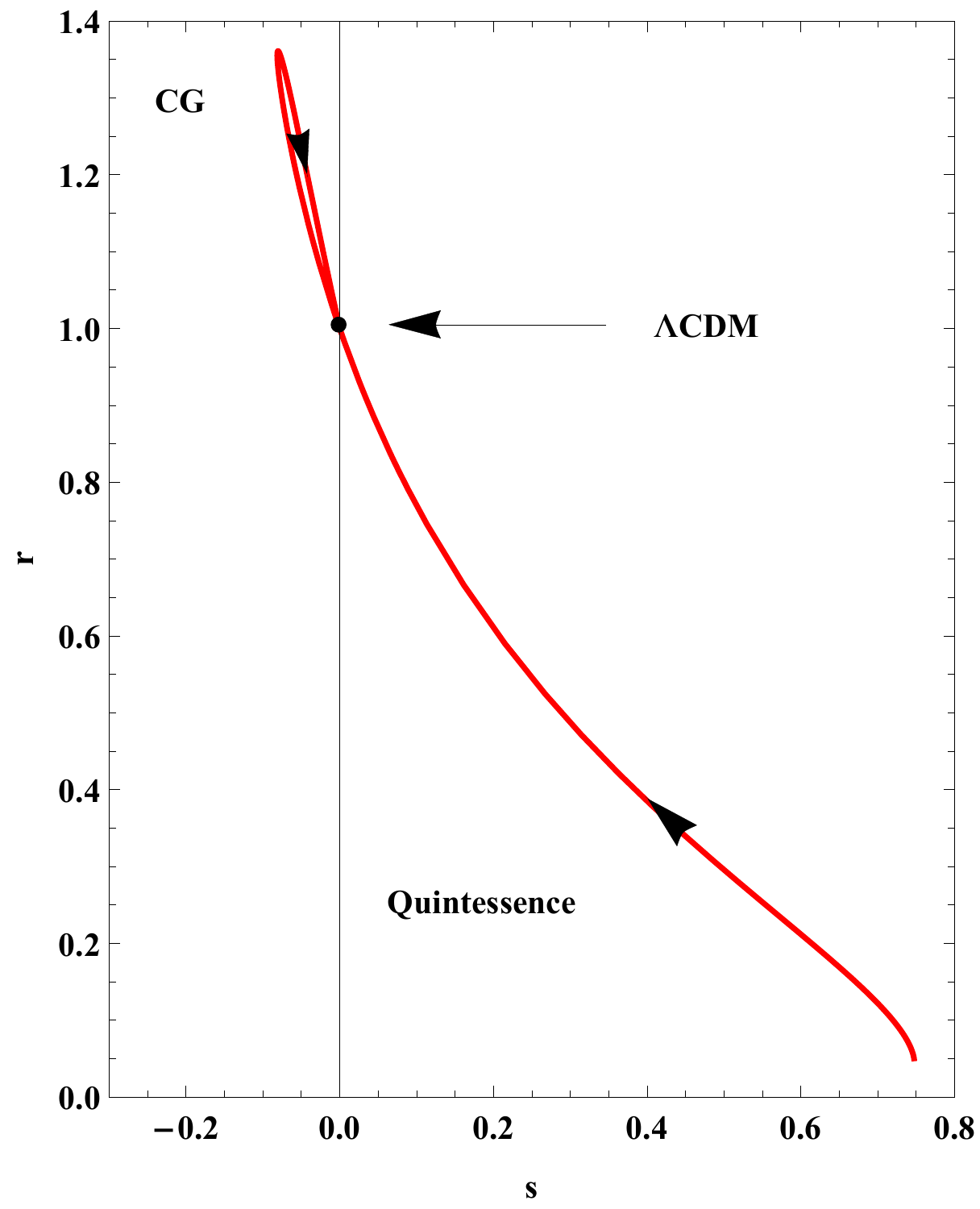}}
\caption{The graphical behavior of the statefinder parameters with the
constraint values from $H(z)$+Pantheon+BAO datasets.}
\label{fig_rs}
\end{figure}

Fig. \ref{fig_rs} represents the $r-s$ plane by considering parameters
constrained by the combined Hubble, Pantheon and BAO datasets. This plot
shows that our model initially approaches the quintessence model ($r<1,s>0$%
). In the later epoch, it reaches the CG model ($r>1,s<0$) and finally it
reaches to the fixed point of $\Lambda $CDM model ($r=1,s=0$) of the
Universe.

\begin{figure}[]
\centerline{\includegraphics[scale=0.7]{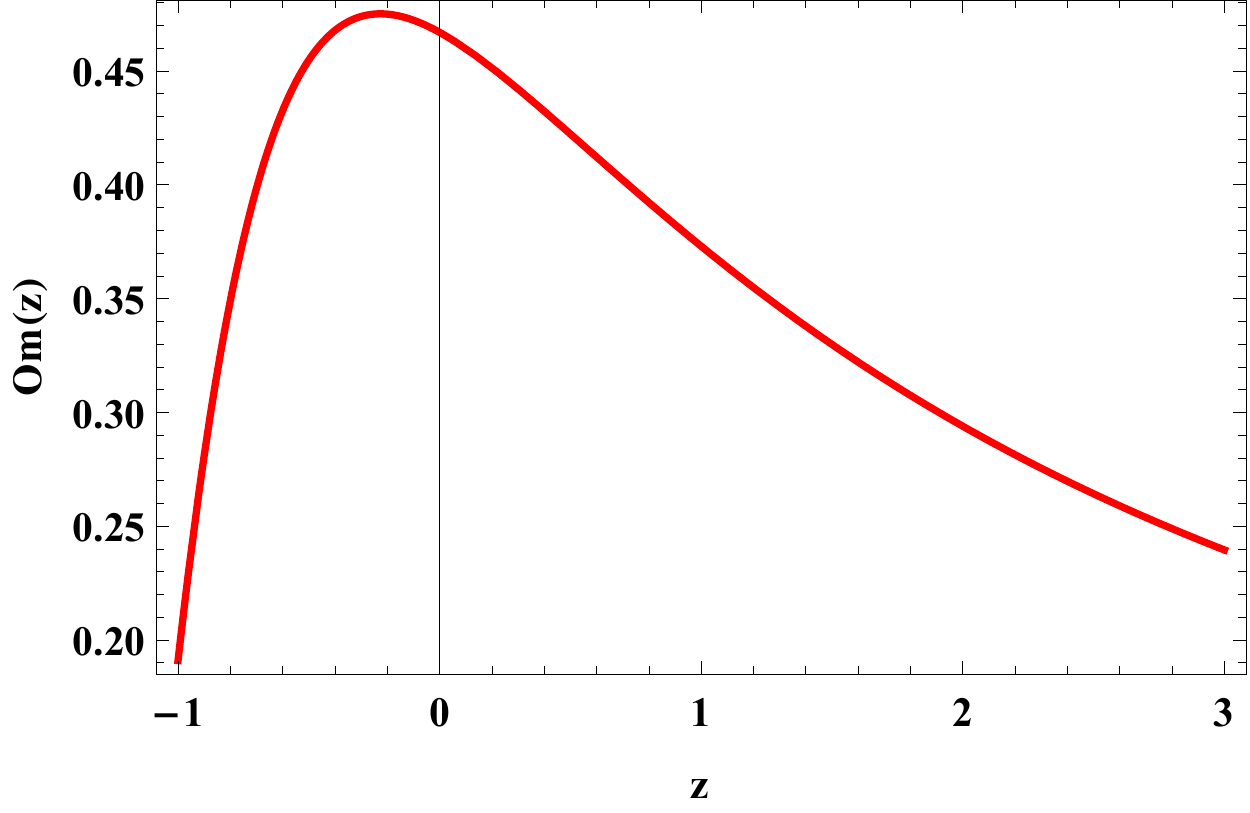}}
\caption{The graphical behavior of the Om diagnostic parameter with the
constraint values from $H(z)$+Pantheon+BAO datasets.}
\label{fig_Om}
\end{figure}

\subsection{Om diagnostics}

The Om diagnostic is another very useful tool to classify the different
cosmological models of DE created from the Hubble parameter \cite{Sahni1}.
It is the simplest diagnostic because it is a function of the Hubble
parameter i.e. it uses only the first-order derivative of the scale factor
of the Universe. In a spatially flat Universe, the Om diagnostic is defined
as 
\begin{equation}
Om\left( z\right) =\frac{E^{2}\left( z\right) -1}{\left( 1+z\right) ^{3}-1},
\end{equation}%
where $E\left( z\right) =\frac{H\left( z\right) }{H_{0}}$ and $H_{0}$ is the
current Hubble constant. This tool allows us to know the dynamical nature of
DE models from the slope of $Om(z)$, for a negative slope, the model behaves
as quintessence while a positive slope represents a phantom behavior of the
model. Lastly, the constant behavior of $Om(z)$ refers to the $\Lambda $CDM
model. The Om diagnostic parameter for our model is 
\begin{equation}
Om\left( z\right) =\frac{\alpha +\beta +\beta z}{z^{2}+3z+3}.
\end{equation}

From Fig. \ref{fig_Om}, it is clear that the Om diagnostic parameter
corresponding to the values of the model parameters constrained by the
combined Hubble, Pantheon and BAO datasets has a negative slope at first,
which indicates the quintessence type behavior, while in the future it
becomes a positive slope which indicates the phantom scenario.

\section{Cosmological $f\left( Q\right) $ model}

\label{sec4}

In this section, we are going to discuss a cosmological models in $f(Q)$
gravity using a new parametrization of the Hubble parameter proposed in the
previous section with the values of the model parameters constrained by the
combined Hubble, Pantheon and BAO datasets. At this stage, the equation of
state (EoS) parameter is used to classify the different phases in the
expansion of the Universe i.e. from the decelerating phase to the
accelerating phase and is defined as $\omega =\frac{p}{\rho }$, where $p$ is
the isotropic pressure and $\rho $ is the energy density of the Universe.
The simplest candidate for DE in GR is the cosmological constant $\Lambda $,
for which $\omega _{\Lambda }=-1$. The value $\omega <-\frac{1}{3}$ is
required for a cosmic acceleration. For other dynamical models of DE such as
quintessence, $-1<\omega <-\frac{1}{3}$ and phantom regime, $\omega <-1$.

For our proposed parametrization of Hubble parameter, we assume a power-law
functional form of non-metricity i.e. \cite{Mandal1},%
\begin{equation}
f\left( Q\right) =Q+mQ^{n},
\end{equation}%
where $m$ and $n$ are the free model parameters. For this specific choice of
the function, by using Eqs. (\ref{F22}) and (\ref{F33}), the energy density
of the Universe and the isotropic pressure can be obtained in the form%
\begin{widetext}
\begin{equation}
\rho =\frac{1}{2}\left( m\left( -6^{n}\right) (2n-1)\left( H^{2}\right)
^{n}-6H^{2}\right) ,
\end{equation}
and 
\begin{equation}
p=\frac{2\overset{.}{H}\left( m6^{n}n(2n-1)\left( H^{2}\right)
^{n}+6H^{2}\right) +3H^{2}\left( m6^{n}(2n-1)\left( H^{2}\right)
^{n}+6H^{2}\right) }{6H^{2}}.
\end{equation}
\end{widetext}

\begin{figure}[]
\includegraphics[scale=0.67]{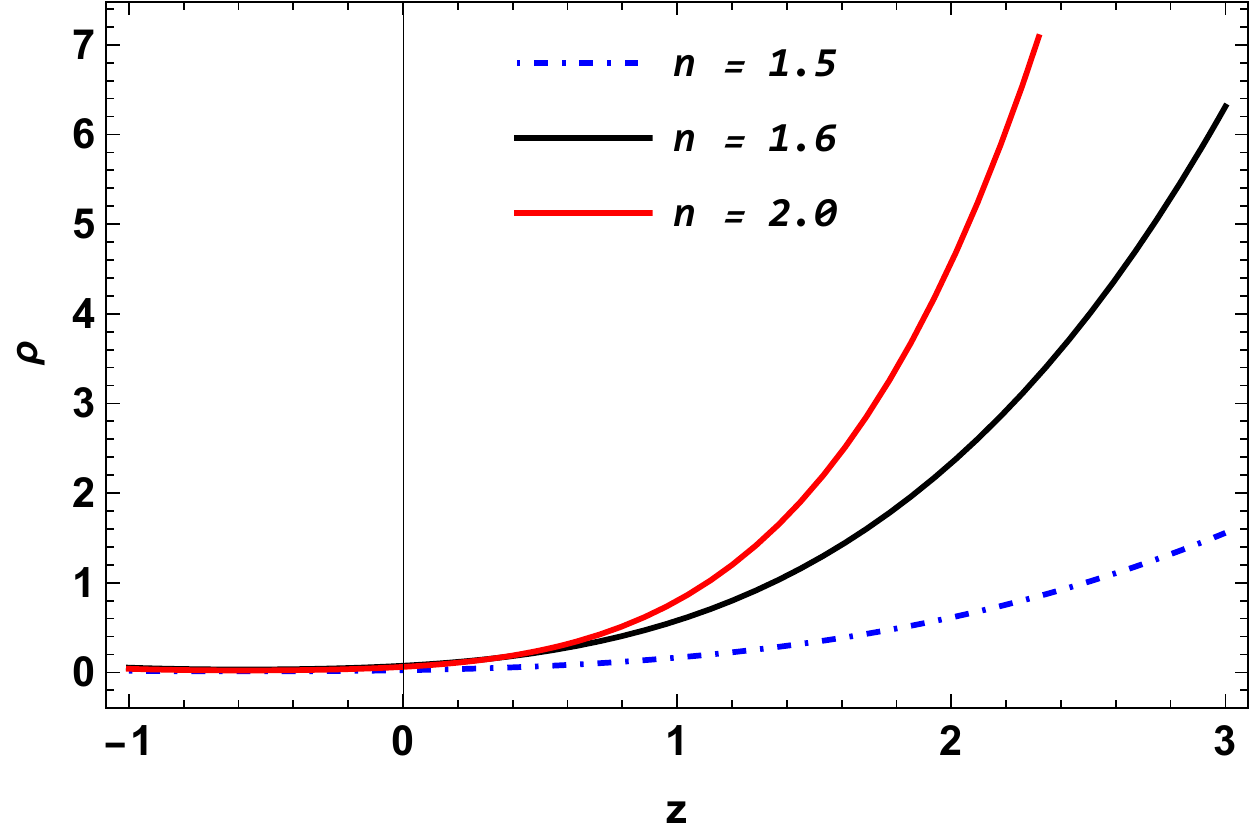}
\caption{The graphical behavior of the energy density with the constraint
values from $H(z)$+Pantheon+BAO datasets and different values for $n$.}
\label{fig_rho2}
\end{figure}

\begin{figure}[]
\includegraphics[scale=0.69]{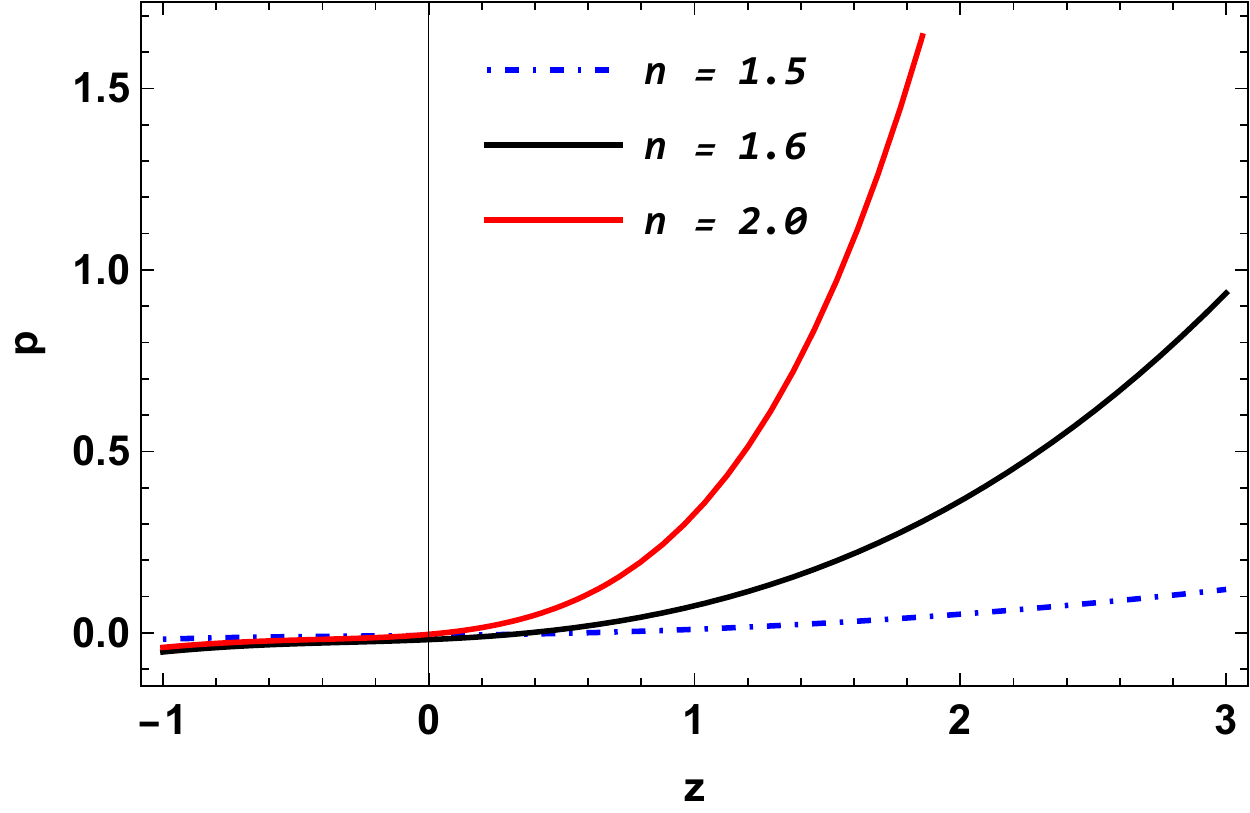}
\caption{The graphical behavior of the isotropic pressure with the
constraint values from $H(z)$+Pantheon+BAO datasets and different values for 
$n$.}
\label{fig_p2}
\end{figure}

\begin{figure}[]
\includegraphics[scale=0.7]{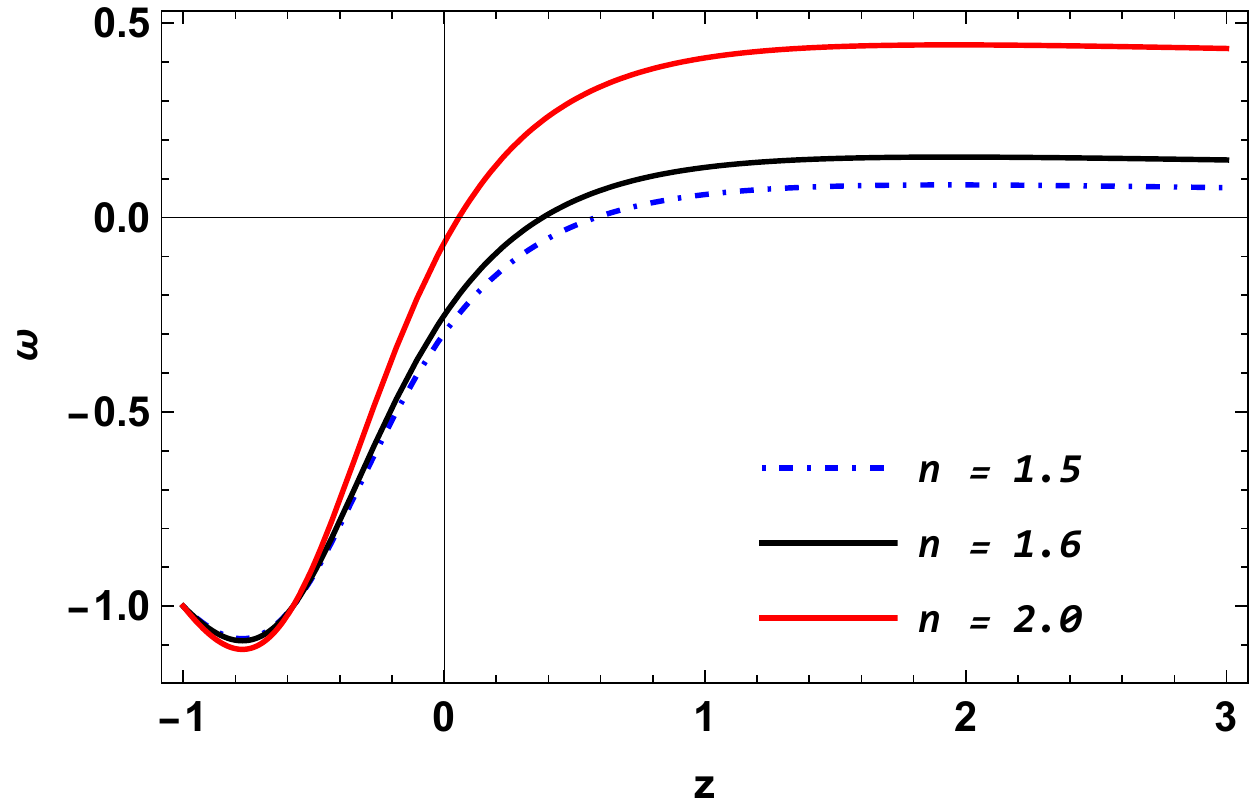}
\caption{The graphical behavior of the EoS parameter with the constraint
values from $H(z)$+Pantheon+BAO datasets and different values for $n$.}
\label{fig_EoS2}
\end{figure}

From Fig. \ref{fig_rho2}, we can observe that the energy density of the
Universe is an increasing function of redshift $z$\ (or a decreasing
function of cosmic time $t$) and remains positive as the Universe expands
for all the three values of $n$. It begins with a positive value and
gradually decreases to zero, as expected. Also, Fig. \ref{fig_p2} shows that
the isotropic pressure is an increasing function of redshift for all the
three values of $n$, which starts with positive values at early times, i.e.
at large $z$,\ and in the late and present time, the pressure becomes
negative with small values close to zero. According to the
observations, the negative pressure is caused by exotic matter such as dark
energy in the context of accelerated expansion of the Universe. 
Fig. \ref{fig_EoS2} depicts the behavior of the EoS parameter vs redshift
using the power-law functional form of non-metricity for various values of $%
n$. It is clear that the model begins from a matter-dominated era ($%
\omega =0$) in early time, traverses the quintessence model ($%
-1<\omega <-\frac{1}{3}$) in the present, and then approaches the $%
\Lambda $CDM region ($\omega =-1$) at $z=-1$. For 
$n=1.5$, the EoS parameter exhibits quintessence-like behavior at the
current epoch, thus both $q$\ and $\omega $\ give the same behaviour.
Further, the present value of the EoS parameter corresponding to the
combined Hubble, Pantheon, and BAO datasets and for different values of $n$\
supports an accelerating phase in this scenario. Finally, it is possible to
see that the behavior of the EoS parameter in our model corresponds to the
models presented in the literature \cite{Koussour1, Koussour2, Mandal3}.

\section{Conclusions}

\label{sec5}

In this paper, we investigated the accelerated expansion of the Universe in
the framework of $f\left( Q\right) $ gravity theory in which the
non-metricity scalar $Q$ describes the gravitational interaction. To find
the exact solutions to the field equations in the FLRW Universe, we proposed
a new parametrization of the Hubble parameter, specifically, $H\left(
z\right) =H_{0}\left[ \left( 1-\alpha \right) +\left( 1+z\right) \left(
\alpha +\beta z\right) \right] ^{\frac{1}{2}}$ where $\alpha $ and $\beta $
are free model parameters, $H_{0}$ represents the present value of the
Hubble parameter. Further, we obtained the best fit values of the model
parameters by using the combined Hubble $H\left( z\right) $, Pantheon and
BAO datasets as $\alpha =0.191_{-0.093}^{+0.093}$, $\beta
=1.21_{-0.13}^{+0.13}$. In addition, we have investigated the behavior of
deceleration parameter, statefinder analysis and Om diagnostic parameter for
the constrained values of model parameters. The evolution of the
deceleration parameter in Fig. \ref{fig_q} indicates that our cosmological
contains a transition from decelerated to accelerated phase. The transition
redshift corresponding to the values of the model parameters constrained by
the combined Hubble, Pantheon and BAO datasets is $z_{t}=0.6857$. Moreover,
the present value of the deceleration parameter is $q_{0}=-0.3092$. Further,
Fig \ref{fig_rs} represents the $r-s$\ evolution trajectories of the model
which initially approaches the quintessence model. In the later epoch, it
reaches the CG model and finally it reaches to the fixed point of $\Lambda $%
CDM model of the Universe. The Om diagnostic parameter has a negative slope
at first, which indicates the quintessence type behavior, while in the
future it becomes a positive slope which indicates the phantom scenario.
Next, to discuss the behavior of other cosmological parameters, we
considered a $f(Q)$ model of the non-metricity scalar, specifically, $%
f\left( Q\right) =Q+mQ^{n}$, where $m$ and $n$ are free parameters. From
Fig. \ref{fig_rho2} we observed that the energy density is positive values
and increasing function of redshift. This represents the expansion of the
Universe. The variation of the isotropic pressure is presented in Fig. \ref%
{fig_p2}. From the figure, we observed that the isotropic pressure
is negative at present and later times. From EoS parameter (see Fig. \ref%
{fig_EoS2}) we observed that $f(Q)$\ model exhibits quintessence-like
behavior at the current epoch. Finally, we conclude that the model supports
the current accelerating Universe.

\textbf{Data availability} There are no new data associated with this
article.

\textbf{Declaration of competing interest} The authors declare that they
have no known competing financial interests or personal relationships that
could have appeared to influence the work reported in this paper.\newline

\acknowledgments 
S. K. J. Pacif \& PKS thank the Inter University Centre for Astronomy and
Astrophysics (IUCAA) for hospitality and facility, where a part of the work
has been carried out during a visit. We are very much grateful to the
honorable referee and to the editor for the illuminating suggestions that
have significantly improved our work in terms of research quality, and
presentation.


\end{document}